\documentclass[reprint,prl,showpacs]{revtex4-1}

\usepackage{graphicx}
\usepackage{amsmath}
\usepackage{hyperref}

\begin{document}

\title{Coherent control of population transfer between vibrational states in an optical lattice via two-path quantum interference}
\author{Chao Zhuang}
\email{czhuang@physics.utoronto.ca}
\author{Christopher R. Paul}
\author{Xiaoxian Liu}
\author{Samansa Maneshi}
\author{Luciano S. Cruz}
\altaffiliation[Current address: ]{Centro de Ci\^encias Naturais e Humanas, Universidade Federal do ABC, Rua Santa Ad\'elia, 166, 09210-170, Santo Andr\'e, S\~ao Paulo, Brazil.}
\author{Aephraim M. Steinberg}
\affiliation{Centre for Quantum Information and Quantum Control and Institute for Optical Sciences, Department of Physics, University of Toronto, Toronto, ON, M5S 1A7, Canada}
\date{\today}

\begin{abstract}
We demonstrate coherent control of population transfer between vibrational states in an optical lattice by using interference between a one-phonon transition at $2\omega$ and a two-phonon transition at $\omega$. The $\omega$ and $2\omega$ transitions are driven by phase- and amplitude-modulation of the lattice laser beams, respectively. By varying the relative phase between these two pathways, we control the branching ratio of transitions to the first excited state and to the higher states. Our best result shows an improvement of the branching ratio by a factor of 3.5$\pm$0.7. Such quantum control techniques may find broad application in suppressing leakage errors in a variety of quantum information architectures.
\end{abstract}

\maketitle

Recent interest in building quantum information processing devices relies on coherent manipulation of quantum two-level systems (qubits). In reality, a qubit is typically a subspace of a larger Hilbert space, meaning that one possible error is leakage to the states outside the computational subspace. This kind of error is different from other kinds of errors, such as fluctuations in control parameters and coupling to the environment\cite{N1367-2630-9-10-384}. Leakage error has been a major concern in many quantum information processing devices, including neutral atoms in optical lattices\cite{ExchangeGate,*PhysRevA.70.012306,PhysRevA.72.052318,*PhysRevA.77.052309}, superconducting qubits\cite{PhysRevLett.83.5385,*PhysRevA.82.040305,PhysRevB.79.060507,*PhysRevA.81.012306,*PhysRevLett.103.110501}, trapped ions\cite{PhysRevA.75.042329,*FaulttolerantIon}, and cavity QED\cite{PhysRevA.67.032305,*PhysRevA.72.032333}.

Here, we present an experiment demonstrating that coherent control techniques\cite{Shapiro2006195,*HerschelRabitz05052000} can be used to suppress leakage errors in quantum information systems, which opens up a new area of application for this kind of techniques. The coherent control scheme we use is analogous to the one-photon vs. two-photon interference scheme\cite{PhysRevB.39.3435}. In that scheme, control is achieved by coherently driving a state with two phase-coherent quantum pathways to the same final state, in which one transition pathway is absorption of a photon at a frequency of $2\omega$ and the other is absorption of two photons at a frequency of $\omega$. The total transition amplitude is the coherent sum of the amplitudes for these two processes, allowing the final-state probability to be controlled by varying the relative phase of the $\omega$ and $2\omega$ transitions. This concept has been applied to coherently control the photoionization of rubidium atoms\cite{PhysRevLett.69.2353}, photocurrents in semiconductors\cite{PhysRevLett.74.3596,*PhysRevLett.78.306} and graphene\cite{DongSun2010}, and photodissociation of molecules\cite{PhysRevLett.74.4799,*PhysRevLett.92.113002}, to name a few proof-of-principle examples. There have also been proposals for using this technique to study the quantum-to-classical transition\cite{PhysRevA.80.053402}, to control photocurrents in carbon nanotubes\cite{PhysRevB.61.7669} and molecular wires\cite{PhysRevB.69.195308}, as well as to control the populations of different electronic states in semiconductor quantum wells\cite{PhysRevB.73.233305} and molecular wires\cite{PhysRevLett.99.126802}. In our work, we extend the application of this technique into the domain of quantum information by demonstrating an analogous scheme based on one-phonon vs. two-phonon interference in a two-vibrational-state system. By using the coherent control technique, we succeed in suppressing leakage during coherent population transfer between the two vibrational states. Our method of controlling vibrational states coherently through different phonon excitations is applicable to situations where phonon excitation is the major source of leakage error, such as the SWAP gate experiments with neutral atoms in optical lattices\cite{ExchangeGate,*PhysRevA.70.012306}. The Hamiltonian of the system we study has exactly the same form as the one for superconducting qubits\cite{PhysRevLett.83.5385,*PhysRevA.82.040305}, suggesting that this method may be useful for suppressing leakage errors in those systems as well.

\begin{figure}
  \includegraphics[width=\columnwidth]{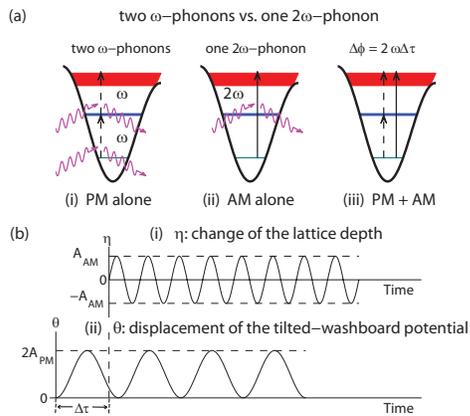}
  \caption{\label{figOneVsTwo} (a) An atom in the ground state can leak out of the qubit space by absorbing either (i) two phonons at $\omega$ (dashed arrows) or (ii) one phonon at $2\omega$ (solid arrow). When the two transitions occur together (iii), the probability of leakage depends on the relative phase between these two transitions. (b) Experimentally, the $\omega$-phonon-transition is realized by phase modulation (PM) of one lattice beam at $\omega$ (ii). The $2\omega$-phonon-transition is realized by amplitude modulation (AM) of the other lattice beam (i). The relative phase $\Delta\phi$ between the two transitions is controlled by the difference $\Delta\tau$ between the initial times when PM and AM are applied, through a relationship of $\Delta\phi=2\omega\Delta\tau$.}
\end{figure}
The two-level system we use is made up of the lowest two vibrational states of an atom trapped in each potential well of an optical lattice. An optical lattice is a periodic potential for atoms, formed by the interference pattern of two laser beams\cite{OpticalLattice}. Because our one-dimensional optical lattice is in the vertical direction, the atoms are actually trapped in a tilted-washboard potential, which is the sum of the periodic lattice potential and the linear potential due to gravity. An atom in the tilted-washboard potential possesses quasi-bound vibrational states, known as Wannier-Stark states\cite{Gluck2002WannierStark,PhysRev.117.432}. We use a shallow tilted-washboard potential which only has two long-lived Wannier-Stark states centered on each potential well. Treating these two states as a qubit, we consider any coupling into the higher excited vibrational states as leakage. To rotate this qubit, we coherently transfer population from the ground state to the excited state by a one-phonon excitation. The one-phonon excitation is experimentally realized by phase modulating (PM) one lattice laser beam at $\omega$ with an acousto-optic modulator (AOM)\cite{PhysRevA.58.R2648,*Phillips2002}, where $\omega$ is the resonance frequency between the ground and first excited states and is measured to be $2\pi\times(4.99\pm0.01)kHz$ for this experiment\cite{PhysRevA.77.022303}. The PM creates a series of sidebands at $\omega_{L}\pm q\omega$, where $\omega_{L}$ is the laser frequency and $q$ is an integer. A Raman transition involving one photon at $\omega_{L}$ and another photon at $\omega_{L}\pm \omega$ can couple two vibrational states with energy separation of $\hbar\omega$. We refer to this creation of a vibrational excitation via absorption of a single modulation quantum at $\omega$ as a one-phonon excitation at $\omega$. An atom in the ground state can absorb one phonon at $\omega$ and be transferred into the excited state, but it can also absorb two phonons at $\omega$ and leak out of the qubit space, as shown in Fig. \ref{figOneVsTwo}(a)(i). To mitigate this leakage, we introduce a second pathway of excitation, a one-phonon excitation at $2\omega$, as shown in Fig. \ref{figOneVsTwo}(a)(ii). Experimentally, we amplitude modulate (AM) the other lattice laser beam at $2\omega$ with an AOM, which creates two sidebands of $\omega_{L}\pm 2\omega$. Similarly, a Raman transition involving one photon at $\omega_{L}$ and another photon at $\omega_{L}\pm 2\omega$ can couple two vibrational states with energy separation of $2\hbar\omega$. We refer to this creation of a vibrational excitation via absorption of a single modulation quantum at $2\omega$ as a one-phonon excitation at $2\omega$. When the two-phonon transition at $\omega$ and the one-phonon transition at $2\omega$ are both driven, as shown in Fig. \ref{figOneVsTwo}(a)(iii), the two pathways interfere, such that the leakage probability depends on their relative phase. By adjusting the pathways to have equal but opposite amplitudes, one could in principle suppress the leakage completely.

\begin{figure}
  \includegraphics[width=\columnwidth]{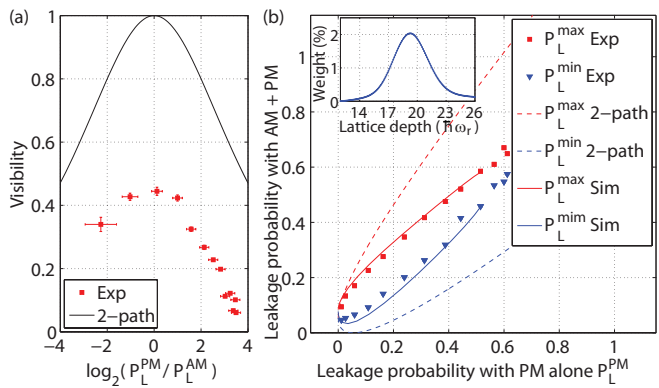}
  \caption{\label{figVisibility} (a) Visibility $(P_{L}^{max}-P_{L}^{min})/(P_{L}^{max}+P_{L}^{min})$ vs. $log_{2}(P_{L}^{PM}/P_{L}^{AM})$; experiment: red dots; idealized 2-path model: solid curve. (b) $P_{L}^{max}$ and $P_{L}^{min}$ vs. $P_{L}^{PM}$ for $P_{L}^{AM}=0.06\pm0.01$. Inset: lattice depth distribution.}
\end{figure}

We use a sample of $^{85}Rb$ atoms laser-cooled to approximately $10\mu K$, with a sufficiently low density (roughly $10^9$ atoms/cm$^3$) that we can neglect interactions between atoms.  Our optical lattice is formed by two 15mW laser beams, red-detuned by 30GHz from the D2 line, which intersect at an angle of $49^{\circ}$, resulting in a lattice spacing of $a = 0.930\mu m$, which is much larger than the 60nm thermal de Broglie wavelength of the atoms. There is therefore vanishing coherence between neighboring wells of the lattice. This lattice has a typical depth of $19\hbar\omega_{r}$, where $\omega_{r} = 2\pi\times h/(8ma^{2}) = 2\pi \times 685Hz$ is the effective recoil frequency and $m$ is the mass of one $^{85}Rb$ atom. Due to the 1.5-mm r.m.s. width of the gaussian lattice beams, the lattice depth is inhomogeneously broadened; the distribution of lattice depths shown in the inset of Fig. \ref{figVisibility}(b) below is measured using the technique in \cite{PhysRevLett.105.193001}. The linear potential of gravity has a value of $2.86\hbar\omega_{r}$ per lattice spacing, and for these parameters there are only two long-lived Wannier-Stark states centered on each lattice well\cite{Gluck2002WannierStark}. By adiabatically lowering the depth of the optical lattice until only one Wannier-Stark state is supported, and then adiabatically increasing it again, we prepare the atoms in the lowest Wannier-Stark state. This same filtering technique\cite{PhysRevA.72.013615} is used to measure the populations of the different vibrational states after excitation.

\begin{figure}
  \includegraphics[width=\columnwidth]{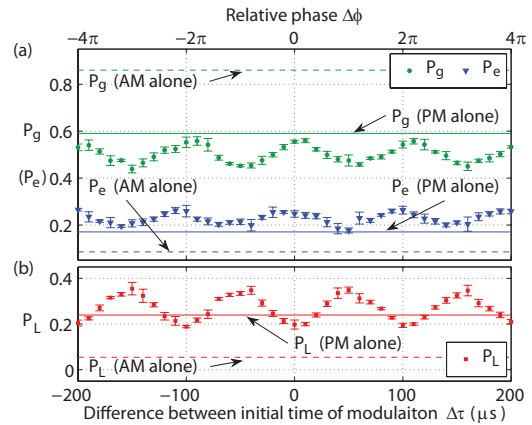}
  \caption{\label{figFringe} The oscillation curves of $P_{g}$ (green dots), $P_{e}$ (blue triangles), and $P_{L}$ (red squares) vs. $\Delta\tau$ when PM and AM are applied together. The control parameters of PM and AM used for these interference fringes are $n=4$, $A_{PM}=8^{\circ}$, $A_{AM}=10\%$.}
\end{figure}
To investigate the interference between the two different transition pathways, we vary the relative phase $\Delta\phi$ between the two-phonon transition at $\omega$ and the one-phonon transition at $2\omega$, while keeping the probability of each transition constant. The probability of each transition depends on the modulation amplitudes $A_{PM}$, $A_{AM}$, and modulation duration $t_{m}$. We always perform the same modulation duration $t_{m}=2n\pi/\omega$ for both PM and AM, where $n$ is an integer. Fig.\ref{figOneVsTwo}(b) shows examples of PM and AM with $n=4$. The modulation duration we use in this experiment is always much smaller than the measured photon scattering time of about $50ms$. The relative phase $\Delta\phi$ between the two transitions depends on the difference $\Delta\tau$ between the initial time of PM and AM, through the relationship $\Delta\phi=2\omega\Delta\tau$. In the first part of the experiment, we measure the probabilities of leaking out of the qubit space $P_{L}$, of being transferred into the excited state $P_{e}$, and of being left in the ground state $P_{g}$, when $\Delta\tau$ is varied and $A_{PM}$, $A_{AM}$, and $n$ are kept constant. Typical oscillation curves of $P_{g}$, $P_{e}$, and $P_{L}$ vs. $\Delta\tau$ measured in the experiment are shown in Fig.\ref{figFringe}, clearly demonstrating interference between the two transition pathways. The constructive and destructive interference conditions for $P_{L}$ are found to be $\Delta\phi=(2l+1)\pi$ and $\Delta\phi=2l\pi$, respectively, where $l$ is an integer. Fig.\ref{figFringe} shows that compared to the leakage $P_{L}^{PM}$ when PM alone is applied, $P_{L}$ is suppressed at the points where destructive interference occurs. The probability of transition into the excited state is increased at the same points. This demonstrates that the leakage due to the two-phonon transition at $\omega$ is suppressed by simultaneously driving the one-phonon transition at $2\omega$ with the appropriate phase.

We further study the dependence of this interference on the leakage probability $P_{L}^{PM}$ by measuring oscillation curves of $P_{L}$ vs. $\Delta\tau$ for different values of the control parameter $A_{PM}$, while $A_{AM}$ and $n$ are kept constant. As all the measured interference fringes show the same constructive and destructive interference conditions, we focus our study on the dependence of the visibility on $P_{L}^{PM}$. We denote the leakage probability when constructive (destructive) interference conditions are met as $P_{L}^{max}$ ($P_{L}^{min}$), which means visibility can be expressed as $(P_{L}^{max}-P_{L}^{min})/(P_{L}^{max}+P_{L}^{min})$. To find $P_{L}^{max}$ and $P_{L}^{min}$, we perform a sinusoidal fit for each interference fringe with the oscillation frequency held constant at $10kHz$. Fig. \ref{figVisibility}(a) shows a plot of visibility vs. $log_{2}(P_{L}^{PM}/P_{L}^{AM})$, where $P_{L}^{AM}$ is the leakage probability when AM alone is applied. We find experimentally (red dots) that the maximum visibility occurs when $P_{L}^{PM} = P_{L}^{AM}$, as would be expected from an idealized two-path interference (2-path) model (solid black line). Such a model assumes the total probability amplitude for leakage is the sum of the amplitudes for the two individual transition pathways, with perfect phase coherence. We would then expect $P_{L}^{max,min} = P_{L}^{PM} + P_{L}^{AM} \pm 2\sqrt{P_{L}^{PM}P_{L}^{AM}}$. Although the experimental result agrees with the 2-path model as to when the maximum visibility should occur, it shows lower visibility.

To understand the discrepancy in visibility between the experimental results and the idealized 2-path model, we consider the Hamiltonian for an atom in the tilted-washboard potential, $H_{0}=p^{2}/(2m)+U_{0}\sin^{2}(\pi x/a)+mgx$, where $U_{0}$ is the optical lattice depth, $g$ is the acceleration due to gravity. By introducing dimensionless parameters $\tilde{x}=\pi x/a$, $\tilde{p}=ap/(\pi\hbar)$, $r=U_{0}/(\hbar\omega_{r})$, and $s=mga/(\hbar\omega_{r})$, we can write the dimensionless time-dependent Hamiltonian in the reference frame which follows the displacement of the potential\cite{Gluck2002WannierStark} in this form:
\begin{equation}
  \mathbf{\tilde{H}_{U}}(t) = \mathbf{\tilde{p}}^{2} + r\sin^{2}\mathbf{\tilde{x}} +
  \frac{s}{\pi}\mathbf{\tilde{x}} -\frac{\ddot{\theta}(t)}{2}\mathbf{\tilde{x}} + r\eta(t)\sin^{2}\mathbf{\tilde{x}},\label{eqnAccHamiltonian}
\end{equation}
where $\theta(t)=A_{PM}(1-\cos\omega t)$ is the displacement of the tilted-washboard potential and $\eta(t)=A_{AM}\sin[2\omega (t-\Delta\tau)]$ is the fractional potential depth modulation, as shown in Fig. \ref{figOneVsTwo}(b). Using the Hamiltonian Eq. (\ref{eqnAccHamiltonian}), we employ a split-operator method to numerically solve the time-dependent Schr\"{o}dinger equation with the ground state as the initial state. Our simulation shows that visibility depends strongly on the lattice depth $r$. Because atoms are trapped in tilted-washboard potentials of different depths, we average the simulation results over the distribution of lattice depths the atoms experience, as shown in the inset of Fig. \ref{figVisibility}(b). To compare the averaged simulation results with the experimental data, we plot $P_{L}^{max}$ ($P_{L}^{min}$) vs. $P_{L}^{PM}$ in Fig. \ref{figVisibility}(b). The simulation results (red and blue solid lines) agree much better with the experimental data (red square and blue triangle) than the idealized 2-path model (red and blue dashed lines). In the simulation, we find that the total leakage is made up of both leakage to other states in the same well and into other wells. Since the destructive interference does not become perfect for the two channels at the same time, visibility is reduced. One would expect the visibility decrease for larger $P_{L}^{PM}$ as well. We believe this inter-well leakage accounts for most of the discrepancy between the experimental data and the 2-path model. This limits the realm of applicability of our control technique to the regime where $P_{L}^{PM}$ is relatively small so the inter-well coupling can be safely ignored.

\begin{figure}
  \includegraphics[width=\columnwidth]{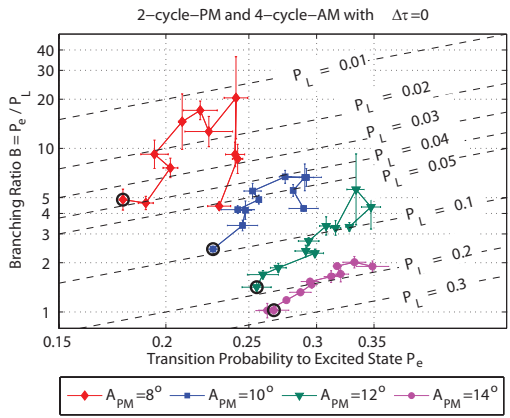}
  \caption{\label{figCompleteSearch} Branching ratio $B=P_{e}/P_{L}$ vs. $P_{e}$ on a log-log plot for $n=2$. For each curve, $P_{L}^{PM}$ is constant by holding $A_{PM}$ constant. The black circle on each curve marks the data when PM alone is applied. The points on the same curve is connected in the order of increasing $A_{AM}$ by increment of $2\%$ with the black circle as the starting point. The dashed lines in the plots are equi-loss lines.}
\end{figure}
In order to test the effectiveness of this coherent control technique, we carry out a search for the best branching ratio in the parameter space of $A_{PM}$, $A_{AM}$, and $n$, while keeping $\Delta\tau=0$ (where the destructive interference occurs). We define the branching ratio as the probability of transition into the excited state divided by the probability of leakage, $B=P_{e}/P_{L}$. If we could completely suppress the leakage, the branching ratio $B$ would go to infinity. It would not be useful to increase the branching ratio by reducing both transition probabilities simultaneously, so we study both figures of merit $B$ and $P_{e}$. Fig.\ref{figCompleteSearch} shows $B$ vs. $P_{e}$ on a log-log graph for a typical set of our experimental data, where $n=2$. For each curve in the graph, we hold $P_{L}^{PM}$ constant and plot the branching ratio in the absence of AM as a black circle for reference. All the other points on the same curve correspond to different values of $P_{L}^{AM}$ (experimentally, different values of $A_{AM}$). As we have learned that the 2-path model does not work perfectly for large values of $P_{L}^{PM}$ and $P_{L}^{AM}$, we expect our coherent control technique to work better for small $n$, $A_{PM}$, and $A_{AM}$. This is confirmed by our experimental results. For $n>5$, we never observe any suppression of the leakage. For $n\leq5$, the branching ratio is always higher for smaller $A_{PM}$ with a given value of $P_{e}$, one can see this in Fig.\ref{figCompleteSearch}. For a given value of $A_{PM}$, increasing the amplitude of AM at first decreases the leakage while simultaneously increasing the excitation probability (and hence the branching ratio). This process reaches an optimum, after which the leakage begins to grow again and the branching ratio declines. The largest enhancement in branching ratio was seen for the smallest value of $A_{PM}$ tested; for the largest values of $A_{PM}$, very little improvement was observed. This agrees with the simulation: when $P_{L}^{AM}$ is too large, we expect the inter-well leakage to become significant, and the degree of leakage suppression is reduced. The largest increase we observed in branching ratio was by a factor of $3.5\pm0.7$, achieved for $n=2$ and $A_{PM}=8^{\circ}$ when $A_{AM}$ was set to $10\%$: the resulting branching ratio was $17\pm2$. Experimental uncertainties prevent us from measuring smaller values of $P_{L}$, but our observations are consistent with the expectation that the enhancement continues to improve for lower drive amplitudes. In summary, we have succeeded in reducing the leakage down to a level limited only by our measurement accuracy.

To conclude, we have experimentally demonstrated a novel coherent control technique for effectively suppressing the leakage error for a two-level system. Our experiment shows that leakage can be suppressed by interference between a two-phonon transition at $\omega$ and a one-phonon transition at $2\omega$ during the coherent population transfer between the ground and excited states of an atom in a tilted-washboard potential. Using this technique, we are able to simultaneously suppress the leakage and increase the transition probability into the excited state. Although our data and analysis show that there are other leakage transitions when the excitation becomes large, it may well be possible to engineer the interference conditions for multiple interference pathways through pulse engineering techniques such as GRAPE \cite{Khaneja2005296,*PhysRevA.85.022306}. The best achieved branching ratio $B=P_{e}/P_{L}$ in our system was $3.5\pm0.7$ times the branching ratio in the absence of coherent control. We believe that similar techniques will prove useful for minimizing leakage in a variety of quantum information architectures, and particularly in those which rely on periodic or washboard potentials, including both optical lattices and superconducting qubits.

We thank Ardavan Darabi for help with the lattice simulations; and Alex Hayat, Matin Hallaji, and Paul J. Godin for useful discussions. We acknowledge financial support from NSERC, QuantumWorks, and CIFAR. LSC acknowledges support from CNPq, Brazil.

\bibliography{mybib}

\end{document}